**Augmenting control arms with Real-World Data for cancer trials: Hybrid control arm methods and considerations**

W. Katherine Tan[a], Brian D. Segal[a*], Melissa D. Curtis[a †], Shrujal S. Baxi[a], William B. Capra[b], Elizabeth Garrett-Mayer[c], Brian P. Hobbs[d], David S. Hong[e], Rebecca A. Hubbard[f], Jiawen Zhu[b], Somnath Sarkar[a], Meghna Samant[a]

[a] Flatiron Health, Inc.; New York, NY 10013

[b] Genentech; South San Francisco, CA, 94080

[c] American Society of Clinical Oncology Center for Research and Analytics (CENTRA); Alexandria, VA 22314

[d] Dell Medical School, University of Texas, Austin, TX 78712

[e] University of Texas M.D. Anderson Cancer Center; Houston, TX 77230

[f] University of Pennsylvania School of Medicine; Philadelphia, PA 19104

* Current affiliation: Indigo Ag; Boston, MA 02129

† Current affiliation: EQRx; Cambridge, MA 02139

**Corresponding author:**

W. Katherine Tan, PhD

Flatiron Health

233 Spring St

New York, NY 10013

Email: ktan@flatiron.com

**Keywords:** Hybrid control arms; external comparator cohorts; real-world data

**Running title:** Hybrid control arms in oncology

**Length details:**

Text, 5,452 words. Abstract, 190. T+F, 8.

**Funding statement:** This study was sponsored by Flatiron Health, Inc., which is an independent subsidiary of the Roche group.

**Author Conflict of Interest Disclosures:** At the time of the study, BDS, MDC, SSB, WKT, SS, MS report employment in Flatiron Health, Inc., and stock ownership in Roche. BPH reports research fundings from Amgen, scientific advisor role and stock ownership in Presagia. RAH reports grant funding from Pfizer. JZ and WBC report employment in Roche/Genentech and stock ownership in Roche. DSH reports research/grant funding from AbbVie, Adaptimmune, Aldi-Norte, Amgen, Astra-Zeneca, Bayer, BMS, Daiichi-Sankyo, Eisai, Fate Therapeutics, Genentech, Genmab, Ignyta, Infinity, Kite, Kyowa, Lilly, LOXO, Merck, MedImmune, Mirati, miRNA, Molecular Templates, Mologen, NCI-CTEP, Novartis, Pfizer, Seattle Genetics, Takeda, and Turning Point Therapeutics; travel and accommodation expenses from Bayer, LOXO, miRNA, Genmab, AACR, ASCO, SITC; consulting or advisory roles with Alpha Insights, Acuta, Amgen, Axiom, Adaptimmune, Baxter, Bayer, COG, Ecor1, Genentech, GLG, Group H, Guidepoint, Infinity, Janssen, Merrimack, Medscape, Numab, Pfizer, Prime Oncology, Seattle

Genetics, Takeda, Trieza Therapeutics, and WebMD; and other ownership interests in Molecular Match, OncoResponse, and Presagia Inc. Other authors: nothing to disclose.

**Acknowledgements:**

We would like to thank Nicole Mahoney, Sheila Nemeth, Khaled Sarsour, and Ashwini Shewade for helpful discussions, and Julia Saiz-Shimosato for editorial assistance.

**Data availability statement:**

The data that support the findings of this study have been originated by Flatiron Health, Inc. These de-identified data may be made available upon request, and are subject to a license agreement with Flatiron Health; interested researchers should contact <DataAccess@flatiron.com> to determine licensing terms.

# Abstract

Randomized controlled trials (RCTs) are the gold standard for assessing drug safety and efficacy. However, RCTs have some drawbacks which have led to the use of single-arm studies to make certain internal drug development and regulatory decisions, particularly in oncology. Hybrid controlled trials with real-world data (RWD), in which the control arm is composed of both trial and real-world patients, have the potential to help address some of the shortcomings of both RCTs and single-arm studies in particular situations, such as when a disease has low prevalence or when the standard of care to be used in the control arm is ineffective or highly toxic and an experimental therapy shows early promise. This paper discusses why it may be beneficial to consider hybrid controlled trials with RWD, what such a design entails, when it may be appropriate, and how to conduct the analyses. We propose a novel two-step borrowing method for the construction of hybrid control arms. We use simulations to demonstrate the operating characteristics of dynamic and static borrowing methods, and highlight the trade-offs and analytic decisions that study teams will need to address when designing a hybrid study.

# 1. Introduction

Randomized controlled trials (RCTs) remain a gold standard for clinical research, but their conduct may become increasingly challenging in oncology.[1] While accelerated regulatory approvals facilitate patients' timely access to effective cancer therapies,[2,3] real-world data (RWD) could foster further research efficiency. Technological advances have boosted data transparency, quality and source traceability, and enabled reliable endpoint analyses that have unlocked the use of RWD sources such as electronic health records (EHRs),[4-7] spurring interest in the use of RWD for drug development and regulatory decisions.[8-13]

RWD can be applied to construct fully external comparator cohorts, but this approach is done without randomization.[14-23] Alternatively, hybrid controlled trial designs that augment RCT control arms with external cohorts (Figure 1) can capitalize on well-developed RWD and still retain the benefits of some randomization. Patients in the external cohort must receive the same treatment regimen as patients in the control arm of the trial, and to the extent possible meet the same eligibility criteria and have comparable treatment and medical histories as patients in the trial.[17,24,25] In the case of EHR-derived data, patients in the external cohort can be contemporaneous to the trial. The external cohort is typically downweighted relative to the randomized control arm at the interim and final analyses based on an *a priori* decision rule to protect against potential biases.

Hybrid control arms are discussed in the FDA guidance for rare diseases.[26,27] In principle, the following are examples where hybrid controlled trials could be beneficial in oncology (Table 1):

- In phase III programs facing challenging timelines (due to long enrollment or follow-up periods) or with secondary interest in low prevalence subgroups,[28] hybrid controlled designs might shorten enrollment periods and shrink overall time horizons, mitigating the risk for premature trial terminations.

- Single-arm phase II trials have traditionally used response rate as the primary endpoint, which may not always be a strong overall survival (OS) surrogate, and can lead to high type I error rates.[29] A hybrid controlled trial with progression-free survival (PFS) or OS as endpoints might provide more reliable evidence for making go/no-go decisions.

- Randomized phase II trials tend to be underpowered to support go/no-go or regulatory decisions[30,31] which can lead to high sign and magnitude error rates.[32] Hybrid controlled designs could increase statistical power and reliability, although the balance between power and bias must be assessed on a case-by-case basis.

For instance, consider IMpassion130, a phase III RCT studying the addition of atezolizumab to nab-paclitaxel to treat metastatic triple-negative breast cancer (mTNBC).[33] This setting has a high unmet therapeutic need, but patients may be averse to randomization to the current standard of care (SOC) of single-agent anthracycline- or taxane-based treatments; whereas immunotherapies such as atezolizumab, have more tolerable side-effects and have shown early promise.[34]

Conversely, hybrid controlled trials would be hard to justify (vs an RCT) in the absence of major barriers to randomization or protracted timelines, or when the unmet need is not high and/or the information about the investigational therapy is limited. Additionally, there may be cases where it is impossible to construct adequate hybrid control arms without unacceptable bias (the external data may not be fit for purpose).

This article discusses methods and considerations for hybrid controlled trials. In particular, we evaluate a few commonly used dynamic borrowing methods, and propose a new frequentist method that, despite its simplicity, performs equally to more complex methods. While these methods can help to protect against unmeasured confounders and other biases, they cannot overcome fundamental differences in patient populations, patterns of care, or endpoint measurements. Any valid RWD application must carefully assess whether or not the data source is fit for purpose, and prospective validation of the data source and trial design may be necessary.[10,12,16,18,19,22,23,35,36]

## 2. Methodology

### 2.1 Components of a hybrid controlled trial

From an analytical perspective, there are four main steps in a hybrid controlled trial beyond what is typically done for an RCT:

1. Select the external cohort based on aligned inclusion/exclusion (I/E) criteria

2. Balance baseline covariates between the trial and external patients

3. Calculate endpoints based on aligned definitions, index dates, and follow-up times across the trial and external patients

4. Implement a borrowing method

The first three steps are common to fully external controls (i.e. single arm trials that use external comparators), and have been described before.[17,19,22,23]

If the external data are deemed fit for purpose in the first, second, and third step (aligning on I/E criteria, balancing covariates, and aligning endpoint definitions, index dates, and follow-up time), then the final step is to implement a borrowing method. There are a number of approaches that can be used to borrow information from external cohorts described in Section 2.2, all of which effectively downweight the external data. Whereas the aim of the first three steps is to account for observed patient and trial characteristics, the aim of this step is to protect against sources of bias that are unknown or cannot otherwise be accounted for.

## 2.2 Existing methods

There are a few commonly used types of borrowing methods,[37,38] including Bayesian approaches such as power prior models, commensurate prior models, meta-analytic predictive (MAP) models, robust MAP models,[39] and hierarchical models, frequentist approaches such as simple test-then-pool procedures, as well as variations of these approaches.

Recently, there have also been proposals for integrating propensity scores into power prior models.[40] However, when patient-level data are available, patient-level matching and weighting (e.g. inverse propensity score weights) may be preferable to the strata-level weights used by Wang et al. (2019), both to retain a clear estimand and possibly to achieve more precise estimates.[40] It would be important to compare these new methods to established patient-level matching and weighting methods before adopting them. For this reason, we currently recommend that patient-level information be used for matching or weighting prior to dynamic borrowing (please see Web Appendix A for details).

Table 2 gives a high-level summary of a few commonly used borrowing methods as well as our proposed two-step procedure described in Section 2.3. These methods can be categorized as being static (the downweighting factor is fixed *a priori*) or dynamic (the downweighting factor is a function of observed outcomes).[37] Each method has a tuning parameter that allows a study team to pre-specify how much they are willing to borrow from the external data.

As noted in Table 2, some methods can determine how much to borrow from external data without accessing data on the experimental patients, whereas other methods do require accessing data on experimental patients. If experimental patient data are used in deciding how much to downweight external patients, the decision could potentially be influenced by the estimated treatment effect, which could raise concerns over the validity of the trial or the need to adjust for multiple hypothesis tests. However, if experimental patient data are never accessed when making this decision, these concerns would not be applicable. Please see Web Appendix B for details on how to measure the amount of information borrowed, as well as Chen et al[41] for an overview of effective sample size.

Table 2 is not exhaustive, and excludes some notable classes of models, such as MAP models.[42] Note that while MAP models are most appropriate for settings where there are multiple external data sources, our context considers an alternative scenario where borrowing is only from one external data source.

## 2.3 Proposed "two-step" borrowing method

**Overview**

We propose a simple "two-step" borrowing procedure as follows:

1. Fit an exponential model to the randomized control and external cohort and estimate the hazard ratio (HR) between the groups. This step does not involve data for the experimental group. Determine the power prior parameter, or weight, as a function of the HR. For example, we set the weight to $w = \exp(-c|\log(HR)|)$ for a pre-specified constant decay factor $c > 0$ (see Figure 2). Similar to the test-then-pool and commensurate prior models, simulations can help inform the choice of $c$.

2. Fit a second exponential model to a dataset containing both the trial and external patients, giving a weight of 1 to all trial patients and a weight of $w$ to all external patients. The second model is used to estimate the treatment effect of the experimental therapy versus the control therapy.

While we used an exponential model in steps 1 and 2 in the simulations of Section 3, it would be straightforward to use a different type of model instead, such as a Weibull or Cox model. Regardless, the same type of model should be used in both steps so that the weights determined in step 1 accurately reflect the amount of residual bias for the model used in step 2.

**Details**

In more detail, let $z_{i,\,experiment}$ and $z_{i,\,external}$ be indicators for whether patient $i$ is in the experimental arm or external cohort, respectively. Also, let $D_{experiment}, D_{control},$ and $D_{external}$ be tabular datasets (all with the same columns and one row per patient) containing data on patients in the experimental arm, control arm, and external cohort, respectively. Suppose that a proportional hazards model is prespecified for the trial analysis.

In step 1 or our proposed approach, the analyst would fit the model

$\lambda(t;\,z_{i,\,external},\,\beta_{external}) = \lambda_0(t) \exp(\beta_{external} z_{i,\,external})$ using the concatenated row bound pooled dataset ($D_{control},\,D_{external}$), where $\lambda(t)$ is the hazard at time $t$ and $\lambda_0(t)$ is the baseline hazard at time $t$. Note that $D_{experiment}$ is not involved in step 1, as data on patients in the experimental arm are not required. The dynamic borrowing weight would then be calculated as $w = \exp(-c|\hat{\beta}_{external}|)$ where the decay factor $c$ is determined prior to analyzing the data through a simulation similar to that in Section 3, in which many values of $c$ are tried in a grid search and one value is selected to achieve the desired operating characteristics.

In step 2 of our proposed approach, the analyst would fit the model $\lambda(t; z_{i, experiment}, \beta_{experiment}) = \lambda_0(t) \exp(\beta_{experiment} z_{i, experiment})$ using the concatenated row bound dataset ($D_{control}$, $D_{external}$, $D_{experiment}$), providing the weight $w$ to all external patients and a weight of 1 to all trial patients. The estimate of the log HR $\hat{\beta}_{experiment}$ and its associated confidence intervals would then be used to determine the effect of the experimental therapy.

As shown in Figure 2, the proposed formula for the weight is equal to 1 if the HR between the randomized and external controls is equal to 1, and decays to 0 as the HR moves away from 1. The larger the value of $c$, the more quickly the weight decays to 0 and the less borrowed information is used. In this way, the weight $w$ is selected dynamically as a function of the data, but the procedure for determining the weight (setting the value of $c$) is specified prior to inspecting and analyzing the data. This is a frequentist analog to the modified power prior, where the weight $w$ is comparable to the power parameter. However, unlike the modified power prior, calculation of $w$ is straightforward and not computationally intensive.

Functional forms other than the one proposed here for determining the weight $w$ may also be useful. We also note that a step function where $w = 1$ at $|\log(HR)| = 0$ and then drops to $w = 0$ at an appropriately chosen value of $|\log(HR)|$ would essentially be a test-then-pool procedure for a point null hypothesis.

## 2.4 Data considerations

**Treatment time period and data collection methods**

The strongest evidence for determining treatment efficacy comes from fully contemporaneous external data including only patients who start therapy after the first patient enrolled in the trial and prior to the last patient enrolled.[26] However, if SOC has remained stable for the particular disease and line setting of the trial, diagnostic practices (e.g. American Joint Committee on Cancer staging criteria) have not changed, and there is no evidence of outcome drift in the time prior to the start of the trial, it may also be possible to include real-world patients who started therapy prior to the first patient enrolled. This could increase the size of the external cohort, which could be particularly relevant for rare diseases. If historical patients are included in the real-world cohort, then additional analytic steps may be needed to ensure that follow-up times are comparable to those in the trial.

In addition to the time period of data collection (historical vs contemporaneous or a mixture of the two), data on the external cohort can be collected either prospectively or retrospectively. Most EHR data is collected retrospectively without the express purpose of supporting clinical trials. However, it is also possible to select patients in a prospectively designed real world study and follow them through the EHR.[43] Prospective intentional data capture makes it possible to obtain additional data to align more precisely with the trial protocol, including I/E criteria, and to better balance key prognostic factors. While retrospective data capture is less burdensome, prospective data capture may allow for better alignment between the randomized and real-world cohorts.

**Assessment of potential benefits**

The assessment and magnitude of potential benefits of a hybrid controlled trial approach is specific to the trial at hand and depends on several factors and assumptions.

Web Appendix C provides a framework for making these assessments with an illustration for a trial similar to IMpassion130,[33] where a potential reduction of the patients randomized to the control arm by half (by effectively accruing enough control patients in the external data source after accounting for downweighting) might have made it possible to reduce the size of the trial by 225 patients, read out the study 4 months early, and enroll patients 2:1 (experimental:control) as opposed to 1:1.

## 3. Simulation study design

To demonstrate how borrowing methods perform, we simulated data resembling a modified IMpassion130 trial[33] if the trial had used a 2:1 randomization ratio instead of a 1:1 ratio, and had been able to effectively borrow half of the control patients from an external data source. The IMPassion130 trial was designed with 88% power, and in the simulations we aimed to achieve the same power but with some events coming from the external data source.

Table 3 shows the simulation scenarios we considered, which cover a range of experimental treatment effects. We also simulated a range of residual bias introduced by the real-world patients after best efforts to select a comparable cohort, balance covariates, and align on endpoints, index dates, and follow-up time. We operationalized this residual bias as the hazard ratio (HR) between the external (real-world) and randomized controls ($HR_{RWD}$).

For each combination of parameters in Table 3, we simulated 1000 datasets and conducted five different analyses: 1) commensurate prior model, 2) test-then-pool procedure with a point null hypothesis, 3) our proposed two-step procedure with an exponential model, 4) power prior model with a fixed power parameter ("static power prior"), and 5) an exponential model to the trial data only (no borrowing) for reference. We then averaged the results over the 1000 datasets to compute the average number of effectively borrowed external events, the type I error rate and power for a one-sided hypothesis test at a 0.025 significance level, the mean squared error and bias of the log hazard ratio comparing experimental and control arms, and the standard deviation of the number of events effectively borrowed (See Web Appendix B for details).

These simulations are intended to reflect the type of assessments that might be done at the design stage of a hybrid controlled trial. They emulate a study design in which the external control arm is fully concurrent, and the final analysis is triggered by the total number of events that have occurred across the trial and external arms (downweighting the events in the external arm based on *a priori* assumptions regarding how much information will be effectively borrowed).

In these simulation settings, the number of external events varies. In particular, we consider the setting where external patients have shorter median survival (ie, relatively more events) than trial controls after steps 1-3 (toward the right of the x-axis in Figure 3), as well as the setting where the external patients have longer median survival (ie, relatively less events) than trial controls after steps 1-3 (toward the left of the x-axis in Figure 3). The range of residual bias on the investigated represents an extreme range of scenarios, some of which may not be relevant in practice. In addition, we consider scenarios where the randomized controls have shorter median survival (ie, relatively more events) than experimental patients ($HR_{Exp} < 1$), as well as the setting where randomized controls have the same median survival (ie, relatively the same number of events) as experimental patients patients ($HR_{Exp} = 1$). Please see Web Appendix B for details on the data generating process, model specifications, and metrics.

As noted above and detailed in Web Appendix B, each method has a tuning parameter. We used grid search to select tuning parameters for each dynamic borrowing method that would result in the lowest type I error while maintaining 88% power for the target $HR_{Exp}$ under no residual bias. While dynamic borrowing methods cap the maximum type I error rate over a range of residual biases, the type I error in any study depends on the actual residual bias, which is unknown in practice. Furthermore, static borrowing methods do not cap the type I error rate, making it difficult to conduct simulations in which the type I error is fixed across both dynamic and static borrowing methods. Consequently, instead of fixing maximum type I error (which is not possible for all methods), we fixed the power under no residual bias and examined potential type I error inflation across a range of residual biases.

## 4. Simulation results

Figure 3 shows the simulation results for the average number of effectively borrowed external events, power (probability of rejecting the null when $HR_{Exp} < 1$), and type I error (probability of rejecting the null when $HR_{Exp} = 1$). Scenarios to the left of the x-axis represent longer median survival in the external controls than randomized controls. With the tuning parameters selected in these simulations and for $HR_{Exp} = 0.78$ and $HR_{RWD} = 1$ (the target scenario under no residual bias), the commensurate prior model has 88.5% power, the test-then-pool procedure has 88.6% power, the two-step approach has 88.5% power, the power prior model with power parameter fixed at 0.6 has 90.2% power, and the reference model that does not borrow any information from external data has 74.1% power.

As seen in Figure 3, the effective number of external events is greatest for the dynamic borrowing methods (commensurate, test-then-pool, and two-step) when the external patients introduce no bias ($HR_{RWD} = 1$), and tapers off as the magnitude of the bias increases. For the static power prior model in this example, the effective number of events is always 60% of the total number of external events. As noted above, there tends to be a greater number of external events as $HR_{RWD}$ increases due to the assumption of equal follow-up time for all groups. Also as noted above, there is a decrease in the number of external events as $HR_{Exp}$ increases to 1 (moving from left column to right of Figure 3) because the hazard in the experimental group becomes similar to that in the control arm, and thus more of the total events occur in the experimental group. This can be seen in the results for the test-then-pool, two-step, and power prior methods. Interestingly, the same trend does not occur for the commensurate prior model.

Regarding power, the left-most panel ($HR_{Exp} = 0.7$) represents an overpowered study, so all methods tend to have near 100% power regardless of residual bias and the number of external events borrowed, except for the static power prior model for which power can be dramatically impacted if there is large residual bias. The second column from the left ($HR_{Exp} = 0.78$) represents the expected experimental treatment effect. The horizontal line is at 88%, which corresponds to the designed power of the IMpassion130 trial. All borrowing methods achieve 88% power when there is no residual bias ($HR_{RWD} = 1$) even though the target number of events was not reached with only trial patients. For all methods, power decreases as fewer events are borrowed. This decrease is more pronounced when the median survival in the external controls is longer than in the randomized controls ($HR_{RWD} < 1$), because the few effectively borrowed events reflect a longer median OS, suggesting that the experimental treatment effect is smaller than it is in truth. The third column from the left ($HR_{Exp}$ 0.85) represents a scenario in which the experimental treatment effect is not as strong as anticipated. Hence the power is shifted downward, but the trends are otherwise similar to the scenario in which $HR_{Exp} = 0.78$. The fourth column from the left ($HR_{Exp} = 1$) represents a scenario in which the experiment treatment has no effect, which is required to assess type I error, as discussed below.

The type I error can be dramatically inflated for the power prior method under large residual bias ($HR_{RWD}$ near 2 when $HR_{Exp} = 1$). However, the dynamic borrowing methods all cap the type I error (max type I error rate of 0.13 for test-then-pool, 0.12 for the commensurate prior model, and 0.097 for the two-step regression), which is shown in Figure 4; these are the same results shown in Figure 3, but excluding the static power prior model and with a different y-axis scale. For these simulations and choice of turning parameters, type I error increases for moderate residual bias ($HR_{RWD}$ near 1.2). However, as the residual bias continues to move away from 1, the models stop borrowing, in turn decreasing type I error. This is a key property of dynamic borrowing methods.[37] We also note that type I error decreases below the nominal rate when the median survival in the external controls is longer than in the randomized controls ($HR_{RWD} < 1$); this is for the same reason that power also decreases in this setting. As the residual bias becomes larger ($HR_{RWD} > 1.2$-$1.3$), and less information is borrowed from the external data, type I error decreases.

By carefully selecting the tuning parameters of the dynamic borrowing methods, we were able to achieve fairly similar performance with the commensurate prior, test-then-pool, and two-step methods. However, in order to obtain 88% in the target scenario with no bias ($HR_{Exp} = 0.78$ and $HR_{RWD} = 1$), the test-then-pool approach incurred the largest max type I error inflation, followed by the commensurate prior model and two-step approach (see Figure 4). The commensurate prior model is more sensitive to residual bias than the two-step approach in these simulations, as seen by the greater type I error inflation and decrease in power, though it might be possible to improve the performance of the commensurate prior model by using a spike-and-slab prior[41] instead of the Half-Cauchy prior we used in these simulations (see Web Appendix B). However, the spike-and-slab prior is also more difficult to tune.

Web Appendix B also shows results for the mean squared error (MSE) and bias of log(HR$_{Exp}$), as well as the standard deviation of the number of external events effectively borrowed.

## 4.1 Assessment of risk and benefits

As noted above, all of these methods have a tuning parameter that can adjust how much is borrowed, which results in different risk/benefit trade-offs.[37] Risk refers to potential inflation of type I error or decrease in power, and benefits refers to potential increases in power, timeline savings, or randomization ratios that allocate more patients to the experimental arm. Figure 5 shows the simulation results for the two-step method with three different tuning parameters, as well as with a model that is fit to the trial data only (no borrowing). As $c$ increases, the weight decays to 0 faster and borrowing is less likely. This results in lower type I error rates and less of a power decrease when there is residual bias, but also lower power when there is no residual bias. Similar trends are observed with the other methods.[37]

By itself, the results shown in Figures 3, 4, and 5 may not provide adequate information to support a study team's decision on which risk/benefit profile they prefer. This decision may depend heavily on the amount of residual bias expected in that setting, and whether the trial is intended for regulatory decision making versus other use cases. To make an informed decision, a study team would need to assess how much bias might be introduced by the external controls after careful cohort selection, covariate balancing, and endpoint, index date, and follow-up alignment (steps 1-3). While it is impossible to know exactly how much residual bias there will be in a particular study, it may be possible to build a body of evidence to suggest likely/plausible scenarios. In particular, by replicating the control arms of recently completed studies (ie, following steps 1-3 and then comparing outcomes with the randomized control) in the same disease area and trial setting, and with the same external data source, it may be possible to develop empirical evidence for how much residual bias might be expected. The operating characteristics of the trial (type I error and power) could then be evaluated accordingly.

For example, Carrigan et al[19] applied steps 1-3 to Flatiron Health's nationwide EHR-derived de-identified database to emulate the control arms of eleven trials in advanced non-small cell lung cancer (aNSCLC), and found that nine trials had a residual bias $HR_{RWD}$ (obtained by exponentiating the 'Difference in ln(HR)' column of Table 1 in that report) between 0.96 and 1.10 for the Overall Survival (OS) endpoint.[19] These authors speculated that this large residual bias was in part due to the enrichment in the trial population of mesenchymal-to-epithelial transition (MET) positive patients, not accounted for in steps 1-3.[19]

Similarly, Tan et al (2021) studied 15 trials across multiple tumor types and found that the majority of trials had $HR_{RWD}$ ranging from 0.66 to 1.09 for the OS endpoint.[23] Such evaluations provide a sense of how much residual bias may be plausible and relevant when selecting the value of tuning parameters and assessing the overall suitability of a hybrid design for a future cancer trials with similar I/E criteria. An evaluation in mTNBC, either based on clinical judgement or an analysis similar to Carrigan et al or Tan et al,[19,23] could help to select the value of tuning parameters and assess overall suitability of a hybrid design for a future mTNBC trial.

# 5. Discussion

Hybrid controlled trials with external RWD have the potential to improve the efficiency of cancer drug development, which could be particularly beneficial in disease settings with low prevalence or long times to event, or for which the SOC has low clinical benefit and/or is very toxic. While we have primarily focused on two-arm designs in this paper, hybrid control arms could also be used in platform trials to support early stage drug development.[44]

Prior to borrowing information from an external source, it is critical to assess whether the external data are fit for purpose. This evaluation involves many factors related to the ability to apply the trial's eligibility criteria to the external dataset (including biomarker and/or genomic information if required), to achieve covariate balance on clinically prognostic characteristics, and to align endpoint definitions, index dates, and follow-up time.[17,22,45]

If the data are deemed fit for purpose, then borrowing methods provide a principled way to protect against unknown or unobservable sources of bias. By using results from control arm replications in similar trials and conducting simulations similar to those in Section 3, drug development teams can assess the amount and impact of residual bias that might remain after careful application of I/E criteria, balancing of covariates, and alignment of endpoints, index dates, and follow-up time. These simulations can also help to assess which borrowing method is preferable for the particular trial at hand, as there is not a clear best model for every scenario. Simulations are also critical for selecting tuning parameters that result in the desired operating characteristics.

While the analytical methods described in this paper help to address potential discrepancies between the trial and external data source, it is always preferable to minimize these discrepancies at the beginning of the study to the extent possible. To this end, we note that treatment patterns in the real world typically follow standard guidelines, such as the National Comprehensive Cancer Network (NCCN) guidelines, and alignment of the trial protocol with these guidelines could reduce the need to rely on analytical methods later in the study to account for differences. Investigations into treatment patterns and patient characteristics in the real-world can also help to inform trial protocols.

As noted above, there is a history of conducting hybrid controlled trials in cancer, though typically with historical trial data as opposed to external RWD.[38,44,46] When bridging historic and current trials, patient populations, endpoint definitions, and assessment timings may be more similar between trials, as compared to RWD. However, RWD may be more recent or collected concurrently to the trial. Using historical data can be problematic when SOCs (including supportive care) or diagnostic methods have evolved over time, or if I/E criteria have become more inclusive.[26,47,48]

For registrational hybrid controlled trials, it will likely be important for the assessment of comparability between randomized and external controls to be conducted by an independent data monitoring committee in a pre-defined manner. Similar to the two-stage design,[49-52] at the interim we recommend that an independent statistician implement the borrowing approach in addition to the weighting or matching. It is typically preferable to have early discussions with regulatory authorities; in the case of the US FDA, we recommend considering study design, operating characteristics of the borrowing methods, and format for submitting RWD,[26,53] possibly through the Complex Innovative Trial Designs Pilot Program.[48] There are also operational features of hybrid control designs that require consideration. In particular, the time at which a sufficient number of events have occurred to make an interim assessment on how much information to borrow from the external cohort may occur when the trial is already or nearly fully enrolled. Furthermore, if the study team had been planning to borrow information from the external data source but the interim assessment shows that it will not be possible, then the trial could potentially be underpowered. In order to mitigate these risks, additional research is needed to develop and assess decision criteria that can be applied early in a trial's enrollment.

In addition, there are many methodological areas for future research to adapt and evaluate borrowing methods for use with external RWD. In particular, it will be important to develop methods for incorporating covariate balancing weights into borrowing methods, including weights to balance post-baseline characteristics and treatment patterns such as differences in treatment duration and subsequent therapies.[17,54] There has already been initial work done in this area,[40] but as described above and in Web Appendix A, we think there may be opportunities for simpler solutions that retain a clear causal estimand. There is also a need to evaluate borrowing methods with simulations that reflect the many nuances of RWD, such as missing data and differential treatment duration, assessment timing, and loss-to-follow-up.

This methodological work is done against the backdrop of a large medical need. Over a million new cancer cases in the United States are projected for 2020[55] but only a small fraction will enroll in a clinical trial.[56] Hybrid controlled trials leverage the overlap between clinical trial protocols and routine care, using valuable patient resources more efficiently to better meet the high unmet needs of patients with cancer. As with any use of RWD, the data sources need to be carefully assessed on a case-by-case basis to ensure the data are fit for purpose, and the operating characteristics of the statistical methods need to be assessed through simulations that mimic the trial at hand. By pairing high-quality external data with rigorous simulations, researchers have the potential to design hybrid controlled trials that better meet the needs of drug development and patients.

## Supplementary material

All R code for the simulations in Sections 3, as well as the potential benefits in Appendix B, are available as online supplementary material.

# Tables

Table 1. Example disease settings and trials for which a hybrid controlled trial may be appropriate to consider. In addition to the considerations outlined in this table, it is critical to weigh the considerations in Section 3.2 to determine whether a hybrid controlled trial is appropriate and whether the external data are fit for purpose.

| Disease setting and representative trials | Low prevalence disease | Long time to events | SOC with low clinical benefit and/or toxic | Comments |
|---|---|---|---|---|
| Metastatic triple negative breast cancer (mTNBC)<br>● IMpassion130 (phase III for atezolizumab) (Schmid et al., 2018) | | | ✔ | ● Median OS < 18 months<br>● Lack of targeted therapies<br>● SOC can be difficult to tolerate (e.g. anthracycline- and taxane-based chemotherapy) |
| Chronic myeloid leukemia (CML)<br>● Phase II for imatinib mesylate (Kantarjian et al. 2002) | | | ✔ | ● Five-year survival for patients diagnosed in 1996-2002, 44.7% (Ries et al. 2006)<br>● SOC at the time (interferon alfa) had limited efficacy and serious side effects |
| Progressive Medullary Thyroid Cancer<br>● EXAM (phase III for cabozantinib) (Eisei et al. 2013) | | | ✔ | ● 10 year survival percentage of 95.6% for local cancers and 40% for metastatic cancers (Roman et al. 2006) |
| | | | | • SOC is ineffective, so placebo was used for control therapy in EXAM. This raises issues as to whether randomization was ethical. |
| Notch activating Adenoid Cystic Carcinoma (ACC)<br>• ACURRACY (clinicaltrials.gov NCT03691207, phase II single-arm)<br>• A future phase III trial | ✔ | | ✔ | • Median OS of ~14 months in general population for ACC (Sharma et al., 2008) (not subset to patients with an activating notch mutation)<br>• Lack of targeted therapy<br>• No established SOC, and common treatments are ineffective and have serious side effects (chemotherapy, surgery, radiation) |
| Adjuvant therapy for early breast cancer<br>• NATALEE (clinicaltrials.gov NCT03701334, phase III for Ribociclib in HR+/HER2-)<br>• APHINITY (phase III for Perjeta + Herceptin in HER2+) (Von Minckwitz et al. 2017) | | ✔ | | • NATALEE is expected to take 7 years to complete.<br>• APHINITY enrolled 4,800 patients to observe 381 invasive disease-free survival events. |
| Pan-tumor NTRK gene fusions<br>• NAVIGATE (clinicaltrials.gov NCT02576431, phase II basket study for larotrectinib)<br>• STARTRK-2 (clinicaltrials.gov NCT02568267, phase II basket study for entrectinib) | ✔ | | May depend on tumor type | • Cohort selection in EHR-derived data may be challenging for basket trials, but might be possible after first gaining experience with each individual tumor type. |

| | | ✔ | | |
|---|---|---|---|---|
| ● A future phase III basket study | | | | |
| First line Diffuse Large B-Cell Lymphoma (DLBCL)<br><br>● GOYA (phase III for Obinutuzumab+CHOP vs Rituximab-CHOP) (Vitolo et al. 2017) | | ✔ | | ● 5 year survival percentage of 62% (Crump et al. 2017)<br>● Rituximab-CHOP has been an established SOC for many years<br>● Approximately one third of patients relapse or are refractory to 1L treatment (Friedberg 2011) |
| Relapsed/Refractory DLBCL<br><br>● ARGO (NCT03422523, phase II for Atezolizumab, Rituximab, Gemcitabine and Oxaliplatin<br>● Potential future studies comparing CAR-NK to CAR-T therapies. This may also be relevant in other disease areas (Liu et al. 2020). | | | ✔ | ● Median OS of 6.3 months for patients whose disease is refractory (best response of progression or stable disease during chemotherapy) or relapses (within 12 months of autologous stem cell transplantation) (Crump et al., 2017) |

Table 2. Common classes of borrowing methods

| Statistical method | Description | Tuning parameter | Pros/cons |
|---|---|---|---|
| *Static* | | | |
| Power prior with fixed power parameter (Chen et al., 2000; Ibrahim et al., 2000) | The contribution of each external patient to the likelihood is weighted by a common "power parameter" between 0 and 1. Typically implemented as a Bayesian model. | Power parameter: Setting it to 1 is equivalent to pooling, and setting it to 0 is equivalent to ignoring external data | **Pro:** Simple and interpretable downweighting factor<br><br>**Con:** Does not cap type I error inflation or decreases in power |
| *Dynamic* | | | |
| Test-then-pool (Viele et al. 2014) | A hypothesis test is done to compare the outcomes of external and trial controls after steps 1-3.<br>● For point null hypotheses, the data are pooled[a] if the null hypothesis of no difference is not rejected, and is ignored otherwise.<br>● For non-equivalency null hypotheses, the external data are pooled if the null is rejected, and is ignored otherwise | For point null hypotheses:<br>● The significance level of the test (smaller alpha makes it more difficult to reject the null, and thus more likely to pool)<br>For non-equivalency null hypotheses:<br>● The significance level of the test (smaller alpha results in wider confidence intervals, making it harder to reject the null and thus less likely to pool)<br>● The equivalency bounds (larger bounds are more likely to contain the | **Pro:**<br>● Simple<br>● Does not require outcome data for experimental group to determine downweighting factor<br><br>**Con:** All or nothing approach, resulting in greater variability and uncertainty about how much information will be borrowed |

| | | confidence interval, thus making it more likely to reject the null and pool) | |
|---|---|---|---|
| Adaptive/modified power prior model (Duan et al. 2006; Neuenschwander et al. 2009) | Similar to the (static) power prior, but the power parameter is given a prior distribution and allowed to be selected based on the data. The power parameter is estimated simultaneously with all other parameters in the model, including the treatment effect. | Hyperpriors on the power parameter | **Pro:** Retains some of the interpretability of the fixed power prior method<br><br>**Con:**<ul><li>Can be difficult to implement in standard software and can be computationally intensive</li><li>Requires outcome data on experimental group to estimate the downweighting factor</li></ul> |
| Frequentist version of modified power prior (See two-step approach in Web Appendix A) | Step 1: A regression model is fit to the external and trial controls to estimate the HR between these two arms. The estimated HR is mapped to a downweighting factor, such that HRs near 1 give a downweighting factor close to 1 and HRs far from 1 give a downweighting factor close to 0.<br><br>Step 2: A second regression model is fit to the pooled external and trial data, giving all external patients the common downweighting factor determined in step 1 and giving all trial patients a | The rate at which the common weights decay to 0 as the HR moves away from 1. For example, the downweighting factor could be defined by the function $w = \exp(c * |log(HR)|)$ for a tuning parameter $c > 0$. Larger values of $c$ result in a faster decay to 0 as the HR moves away from 1. | **Pro:**<ul><li>Simple and interpretable downweighting factor that is chosen dynamically</li><li>Does not require outcome data from experimental group to determine downweighting factor, as the downweighting factor is determined in step 1 and outcome data for the experimental group is not required until step 2</li></ul> |

| | | | |
|---|---|---|---|
| | weight of 1. | | |
| | | | **Con:** Still pending a full evaluation of performance in different settings |
| Commensurate prior model (Hobbs et al. 2011, 2012) | The outcomes in the randomized controls are centered around the outcomes in the external controls. For example, the log hazard rate of the trial controls might be given a normal prior, centered around the log hazard in the external controls and with hyperprior on the precision of the normal prior. | The hyperpriors on the precision of the normal distribution that shrinks the hazard rate in the randomized controls toward the hazard rate in the external controls. The more this precision is pushed toward zero, the less the hazard in the trial controls is shrunk toward the hazard in the external controls and the more the external controls are effectively downweighted. | **Pro:** Dynamic Bayesian borrowing method that is straightforward to implement in standard software<br><br>**Con:**<br>● Downweighting is implicit, so can be more difficult to interpret the amount of borrowed information.<br>● Requires outcome data on experimental group to estimate the downweighting factor |

[a]In this context, pooling refers to combining RWD and trial control data into a single dataset that is then analyzed as though the data were collected together.

Table 3. Simulation setup based on IMpassion130[33].

| Parameter | Values |
| --- | --- |
| Experimental treatment effect: Hazard ratio between experimental and control arms of trial ($HR_{Exp}$) | 0.70 (More effective than expected)<br>0.78 (Target HR, i.e. alternative hypothesis)<br>0.85 (Less effective than expected)<br>1.00 (No treatment effect) |
| Residual bias: Hazard ratio between real-world controls and randomized controls after careful alignment on I/E criteria, covariate balancing, and alignment of endpoints, index dates, and follow-up time ($HR_{RWD}$) (composite bias) | Range from 0.5 to 2 by 0.1 (i.e. 0.5, 0.6,..., 1.9, 2.0):<br>0.5 (Extreme): External patients have longer median time-to-event than randomized controls<br>1 (No bias)<br>2 (Extreme): External patients have shorter median time-to-event than randomized controls |
| Expected downweighting factor for external controls[a] | 0.6 |
| Total number of patients in RCT (control + experimental) | 675 (out of 900 planned in IMpassion130) |
| Number of external patients potentially available to borrow | 375 (resulting in an expected 375 * 0.6 = 225 effectively borrowed external patients) |
| Randomization ratio in trial | 2:1 (experimental:control) |
| Target number of events (control + experimental + downweighted external control) | 655 |
| Percent lost to follow-up in both the trial and external data source | 5% |
| Accrual rate in trial | 34 patients per month |
| Significance level for hypothesis test of experimental treatment effect | 0.025 one-sided |

[a]At the time of study design, the downweighting factor is known with certainty if using a power prior model with fixed power parameter, and is predicted if using a dynamic borrowing method.

# Figure Legends

Figure 1. Example schema for a hybrid controlled trial using external RWD

Figure 2. Simulation results. X-axis values smaller than 1 indicate that external controls have longer median time-to-event than randomized controls after steps 1-3, and x-axis values larger than 1 indicate that external controls have shorter median time-to-event than randomized controls after steps 1-3. In practice, the full range of residual bias shown on the x-axis may not be relevant (see Section 4.4).

Figure 3. Type I error excluding power prior model. These are the same results shown in Figure 2, but excluding the power prior model and with a different y-axis scale. In practice, the full range of residual bias shown on the x-axis may not be relevant (see Section 4.4).

Figure 4. Two-step procedure with different risk/benefit profiles. In practice, the full range of residual bias shown on the x-axis may not be relevant.

Figure 1.

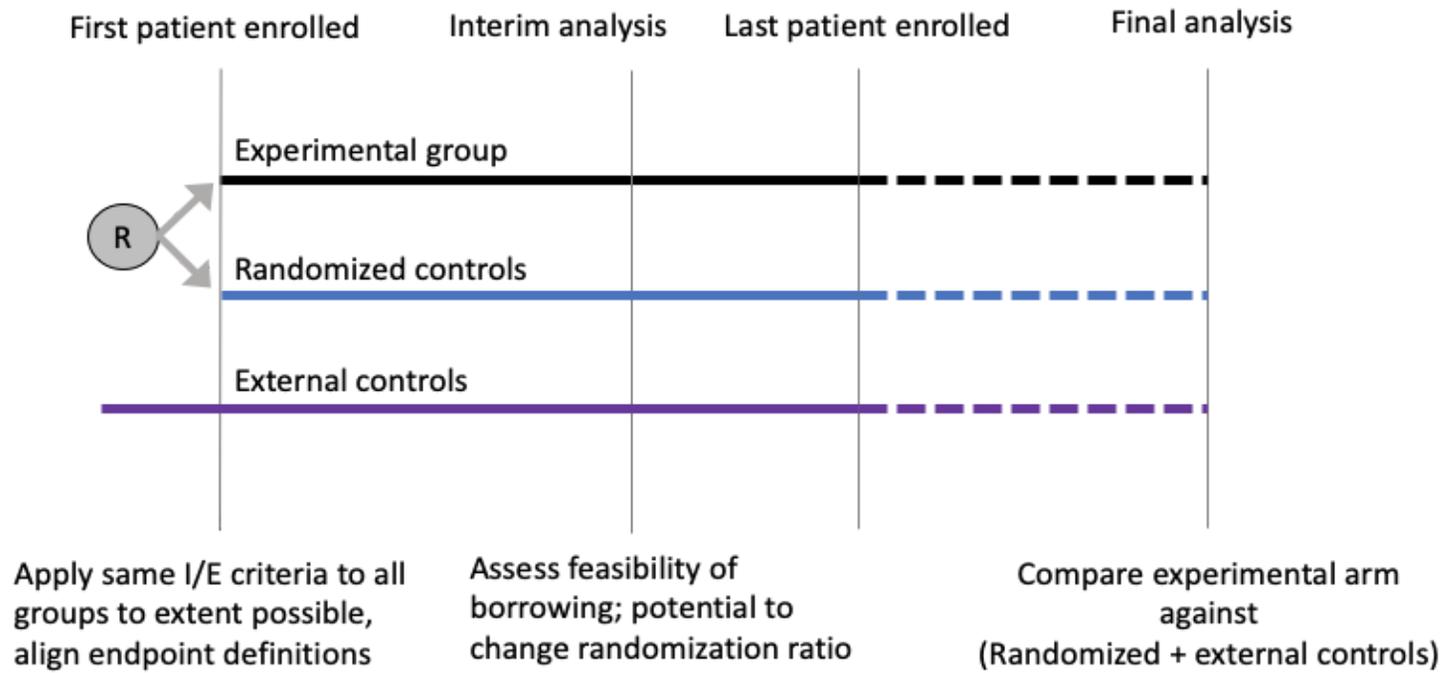

Figure 2.

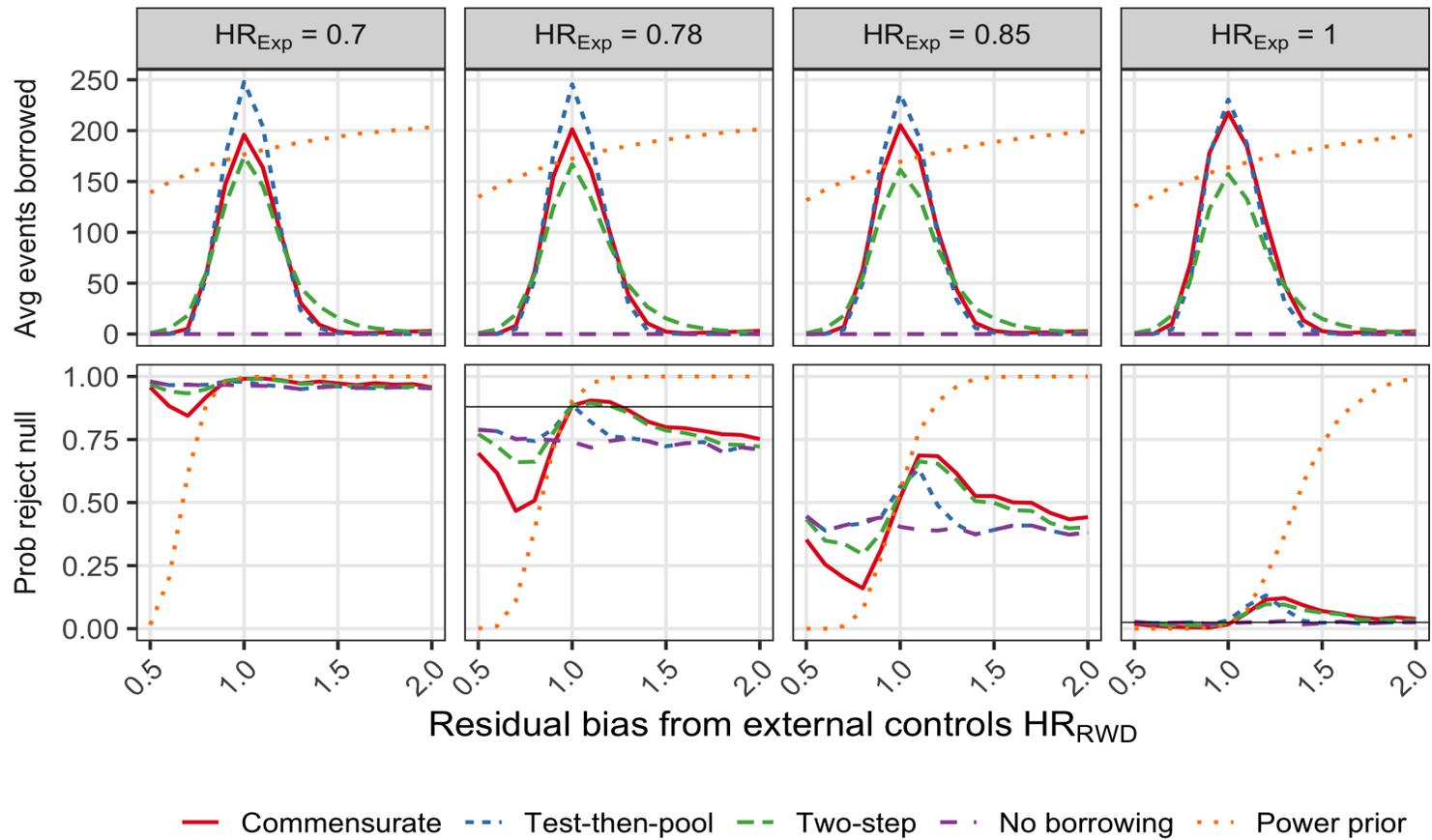

Figure 3.

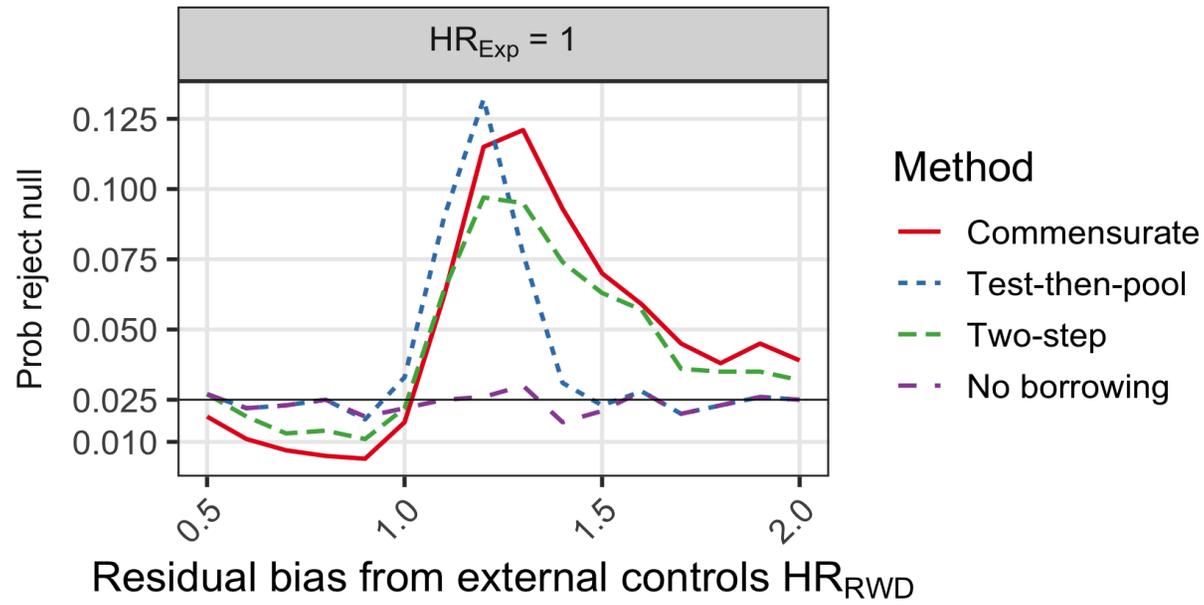

Figure 4.

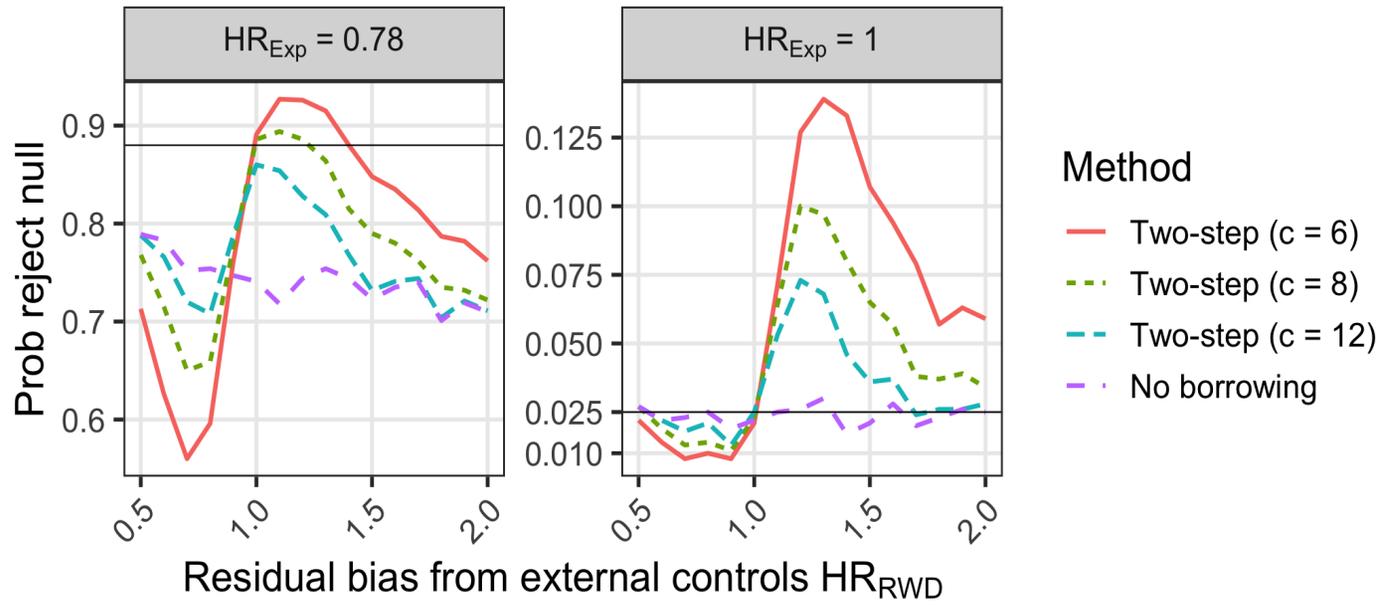

Online supplementary material

Augmenting control arms with Real-World Data for cancer trials: Hybrid control arm methods and considerations

Table of contents

# Appendix A: Using covariate balancing weights in borrowing methods

If a matching method is used to balance covariates, then the matched dataset can be directly used in subsequent borrowing methods. However, it may not always be practical to use a matching method, such as when patient counts are low. In this case, a weighting method might be used to balance covariates, such as inverse propensity weighting. If the estimand of interest is the treatment efficacy among the trial population, then the average treatment effect on the treated (ATT) weights could be formed by assigning weights of 1 to all patients in the trial and inverse odds of not being in the trial to external patients (see Li, Morgan, and Zaslavsky 2018).

It may be possible to use these covariate balancing weights in subsequent borrowing methods. For example, for the test-then-pool procedure, a weighted log-rank test could be conducted to determine whether to pool the data, and the same weights could be used in the second step to estimate the experimental treatment effect. Similarly, in the two-step Frequentist analog to the modified power prior, the regression model fit in the first step could be weighted by the covariate balancing weights, and then each patient's covariate balancing weight could be multiplied by the cohort-level downweighting factor when fitting the regression model in the second step. In the power prior and commensurate prior model, it may be possible to weight each patient's contribution to the likelihood by their covariate balancing weight. For the power prior model, this would result in multiplying each patient's covariate balancing weight by the cohort-level downweighting factor, just as in the two-step approach. Investigating the feasibility and performance of these approaches could be an important area for future work.

Because all external data are non-randomized, it is critical to specify the causal estimand of interest, such as the ATT for efficacy or the average treatment effect (ATE) for effectiveness. If all sources of selection bias and confounding could be accounted for through careful application of steps 1-3 (cohort selection, covariate balancing, and alignment of endpoints, index dates, and follow-up), there would be no need for borrowing methods. An accurate and likely more efficient estimate of the experimental treatment effect could be obtained by following steps 1-3 and then conducting an analysis with the weighted or matched dataset. Of course, it is not possible to know if steps 1-3 have fully accounted for all sources of bias, in which case borrowing methods, particularly dynamic borrowing methods, can help to protect against unknown sources of bias. However, in the case where there is no residual bias, we think the borrowing method should produce a consistent estimator for the causal estimand of interest. This would be the case for a test-then-pool procedure with ATT or ATE weights, and further investigations into this topic for other borrowing methods could be an important area for future research.

## Appendix B: Illustration of potential benefits

### B.1: Overview

In this appendix, we illustrate these potential benefits for a trial similar to IMpassion130 (Schmid et al. 2018) had it been designed with a hybrid control arm and assuming the external data does not introduce any residual bias. These illustrations prioritize the collection of concurrent external data, which adds to the uncertainty in study design but may also provide stronger evidence (US Food and Drug Administration 2020). As noted above, there are also potential risks which must be weighed alongside these benefits when designing a trial.

Table B.1 shows the study design parameters for a trial similar to IMpassion130. The RCT parameter values in Table B.1 are based on the co-primary OS intent-to-treat (ITT) analysis for IMpassion130, and the hybrid design parameter values are intended to demonstrate the potential efficiencies if data could be borrowed from an external source. This illustration is under the ideal scenario in which the study team knows at the start of the trial how many external patients will accrue per month, and how many of these patients they will be able to effectively borrow. In practice, we recommend using a conservative estimate for the effective external accrual rate (after downweighting) to avoid underpowered trials. Please see Web Appendix B.2 for details on how the hybrid design parameters were determined.

We note that the number of external patients effectively borrowed represents the effective amount of information after downweighting. For example, suppose 375 external patients meet the I/E criteria and a weighting method such as inverse probability of treatment weights is used to balance covariates, which estimates weights $w_i$ for external patients $i = 1, ...375$ (the patient-level weights $w_i$ used to balance covariates are distinct from the cohort-level weight used in dynamic borrowing to downweight external patients). Also suppose that a power prior or two-step borrowing approach is used and that the cohort-level downweighting factor is determined to be $0 < a < 1$. Then the number of patients effectively borrowed could be approximated as $a \sum_{i=1}^{375} w_i$ (please see Web Appendix C for additional information). In the case where covariates are perfectly balanced prior to the weighting step such that $w_i = 1$ for $i = 1, ...375$ and the external data are pooled such that $a = 1$, the effective number of external patients would be 375. Alternatively, suppose $\sum_{i=1}^{375} w_i = 300$ and $a = 0.75$. Then the effective number of external patients would be 225.

**Table B.1.** Study design parameters for illustration of potential benefits. Values for the non-hybrid design are based on but not identical to the co-primary OS ITT analysis for IMpassion130 (Schmid et al. 2018). See Web Appendix B.2 for details on how the hybrid design parameters were determined.

| Parameter | Original non-hybrid design | Alternative hybrid design |
|---|---|---|
| Total number of patients in RCT | 900[1] | 675 |
| Total number of external patients effectively borrowed after all weighting steps | N/A | 225 |

---

[1] In the trial, IMpassion130 enrolled 902 patients.

| Accrual rate (assumed linear) | 34[2] patients/month in trial | 34 patients/month in trial[3] 11.3 patients/month in external controls (effectively borrowed after weighting)[4] |
|---|---|---|
| Expected hazard ratio for experimental treatment | 0.78 | 0.78 |
| Expected median OS in control group | 16 months | 16 months in both trial and external controls. This implies no residual bias, which is a key assumption that would need to be carefully evaluated in practice. |

---

[2] The IMpassion130 protocol assumed enrollment of 900 patients would take 26 months. Assuming linear enrollment, this is 34.6 patients / month
[3] This makes the conservative assumption that the enrollment rate stays the same even with a more favorable randomization ratio
[4] In practice, this could mean that for every 2 months in which 11 patients effectively accrue in the external cohort, there is one month in which 12 patients effectively accrue.

| Expected annual loss to follow-up in both the trial and external data source | 5% | 5% |
|---|---|---|
| Randomization ratio (experimental:control) | 1:1 | 2:1 |
| Target number of events | 655[5]:<br>• 310 experimental<br>• 345 control | 655:<br>• 310 experimental<br>• 345 control<br>    ○ 173 trial<br>    ○ 172 external |
| Expected time to finish | 26[6] months: | 20 months |

---

[5] The breakdown of expected events per arm is based on the calculations in Appendix C.1..
[6] In the trial, IMpassion130 finished enrollment in 23 months

| enrollment | | |
| --- | --- | --- |
| Expected time to clinical cut-off date[7] | 53 months | 49 months (assumes no residual bias) |
| Statistical power | 88% | 88% |

Figure B.1a shows the RCT trial design as specified in Table B.1 without any external data, and Figure B.1b shows the hybrid controlled trial design. In both figures, the top panel shows cumulative study enrollment over time, and the bottom panel shows the cumulative number of events that would be expected to occur over time. As can be seen in these figures, the total sample size and number of events for experimental and control patients are the same in the two designs, but the source of the control patients is split evenly in the hybrid trial between the randomized and external arms.

In effect, this hybrid design increases the monthly accrual in the trial by 11-12 patients per month, leading to enrollment completion 6 months early, study read-out 4 months early, 225 fewer patients in the trial, and a 2:1 randomization ratio as opposed to a 1:1 (experimental:control).

---

[7] This assumes that a patient's death is reported immediately after it occurs in both the trial and EHR-derived dataset. This may be appropriate for prospectively collected EHR data, but might not hold for retrospectively collected data.

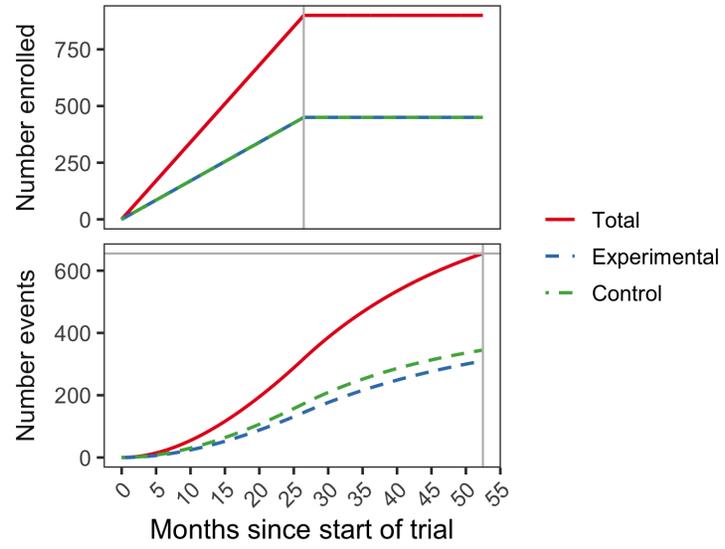

**Figure B.1a.** Excluding external data

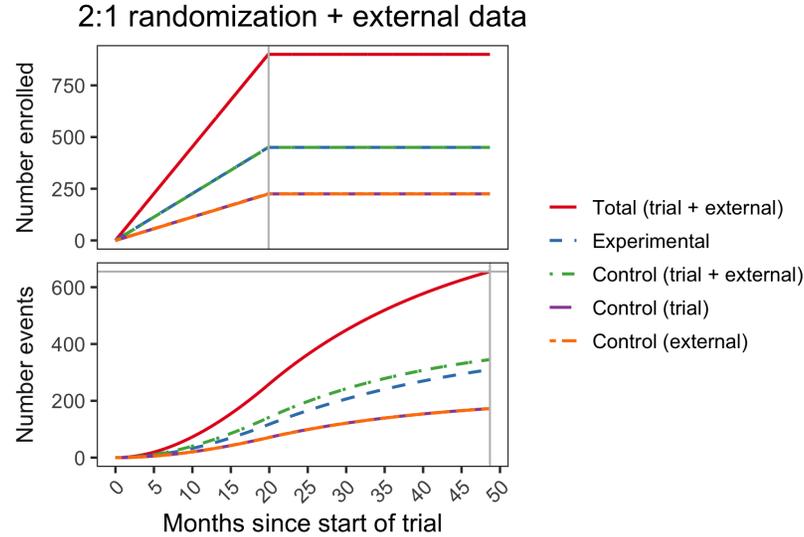

**Figure B.1b.** Including external data

**Figure B.** Trial design with and without external data.

The potential benefits discussed in this section assume that the study team can accurately assess the future accrual rate in the external data prior to beginning the trial. This may be a difficult task, but can be informed by carefully scoping historical accrual rates in the external data among patients meeting the trial's I/E criteria. There is no guarantee that historical trends will continue, so conservative assumptions may be warranted. Previous efforts to replicate the control arm of similar trials using the same external data source would also be important to assess the amount of information that will likely be borrowed after downweighting.

## B.2: Details for illustration of potential benefits

Let $r:1$ be the randomization ratio in the trial (experimental to control), $N_{RWD}(t)$ be the number of external (RWD) controls enrolled on or before time $t$, where $t = 0$ is when the first site activates (first enrolled patient may be after site activation), and let $n_C$ and $n_E$ be the total number of planned control and experimental patients respectively to be enrolled in the trial. Let $s_E$, $s_C$, $s_{RWD}$ be the rate at which patients enroll in the experimental, randomized control, and external (RWD) control arms respectively (e.g. patients / month). The total enrolment rate in the trial is $s = s_E + s_C$.

For fixed total enrollment rates in the trial and external cohort ($s$, and $s_{RWD}$), fixed months of historical external accrual $t_{historic}$, and initial randomization ratio $r:1$ in the trial (experimental:control), we update the randomization ratio $r$ and enrollment rates in the experimental and control arms ($s_E$ and $s_C$) to account for external patients as follows:

1. Set the number of historical external controls to $N_{RWD}(0) = t_{historic} * s_{RWD}$

2. Update the randomization ratio and enrollment rates in the trial to account for historical external controls:

   a. $r \leftarrow n_E / [n_C - N_{RWD}(0)]$

   b. $s_C \leftarrow s * r / (r + 1)$

c. $s_E \leftarrow s / (r + 1)$

3. Update the enrollment rates and randomization ratio in the trial to account for the concurrent external controls such that accrual of the experimental, trial control, and external control arms will finish at the same time:

   a. Set $s_C$ and $s_E$ to the solution to the following system of 2 linear equations (Note: It takes $n_E / s_E$ units of time to complete enrollment in the experimental arm and $[n_C - N_{RWD}(0)] / (s_C + s_{RWD})$ units of time to complete enrollment in the hybrid (trial + external) control arm. Setting these two amounts of time equal to each other, inverting, and splitting the second term gives the second constraint below):

      i. $s_C + s_E = s$

      ii. $s_E (1 / n_E) - s_C (1 / [n_C - N_{RWD}(0)]) = s_{RWD} / [n_C - N_{RWD}(0)]$

   b. Update the randomization ratio to $r \leftarrow s_E / s_C$

We then projected accrual and event times as follows:

1. Create a vector of accrual times assuming linear enrollment:

   a. $u_C = (1/s_C, 2/s_C, ..., [n_C - N_{RWD}(0)]/s_C)$

b. $u_E = (1/s_E, 2/s_E, ..., n_E/s_E)$

c. $u_{RWD} = (-(a-1)/s_{RWD}, -(a-2)/s_{RWD}, ..., 0, 1/s_{RWD}, 2/s_{RWD}, ..., n_{RWD, conc}/s_{RWD})$ where $n_{RWD, conc} = s_{RWD} * (n_E / s_E)$ is the total number of concurrent external controls that will enroll (the first term on the left-hand side is the number of patients accruing per unit time in the external cohort and the second term is the length of the study).

2. Let $F_\lambda$ be an exponential cumulative distribution function with rate λ, let $p_{lost}$ be the proportion lost to follow-up, let $dN_E(t)$, $dN_C(t)$, and $dN_{RWD}(t)$ be the number of patients enrolling at time $t$ in the experimental, randomized control, and external (RWD) cohort respectively, and let $d_E(t)$, $d_C(t)$, and $d_{RWD}(t)$ be the number of events that have occurred by time $t$ in the experimental, randomized control, and external cohort respectively. Using $i$ as a subscript to denote the elements of $u_C$, $u_E$, and $u_{RWD}$, we computed expected number of events at time $t > 0$ as:

   a. Experimental group: $E[d_E(t)] = (1 - p_{lost}) \sum_{i:u_{E,i} \leq t} dN_E(t) F_\lambda(t - u_i)$ where the hazard rate is $\lambda = 0.043 * 0.78$.

   b. Randomized control group: $E[d_C(t)] = (1 - p_{lost}) \sum_{i:u_{C,i} \leq t} dN_C(t) F(t - u_i)$ where the hazard rate is $\lambda = 0.043$.

c. External control group: $E[d_{RWD}(t)] = (1 - p_{lost}) \sum_{i:u_{RWD,i} \leq t} dN_{RWD}(t) F(\min(t - u_i, t))$ where the hazard rate is $\lambda = 0.043$. For each time $t$, the maximum follow-up time is set to $t$ to ensure that historical external patients do not have longer follow-up than what is possible in the trial. This is demonstrated in the Figure B.2 below for a hybrid design with 6 months of historical patients and 11.3 effectively borrowed external patients accruing per month on average. Note that in Figure B.2 the external accrual rate does not change over the course of the study (this is fixed), and no events are recorded in the external cohort prior to the start of the trial. This is due to the maximum follow-up time restriction, which only allows external events to be uncensored if the trial duration is at least as long as the external time-to-event.

From the output of these calculations, we get the updated randomization ratio that takes into account historical and concurrent external patients, as well as the expected time to finish enrollment and to hit the target number of events across the trial and external arms.

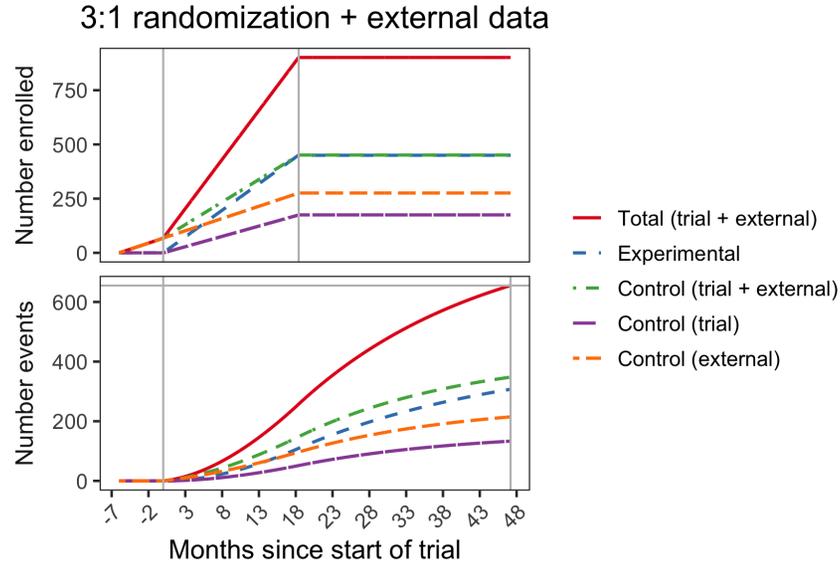

**Figure B.2.** Enrollment and event projections if there were 6 months of historical external data and an accrual rate of 11.3 effectively borrowed external patients per month on average. The randomization ratio is rounded to the nearest whole number.

## Appendix C: Details for simulations

This appendix provides details on the simulations in Section 4.

### C.1 Data generating process

We simulated data for fully concurrent external data in a way that ensures equal follow-up time across the trial and external arms. We did this by fixing the desired randomization ratio at 2:1 and the trial accrual rate at 34 patients per month (assumed fixed over time), and then calculating the number of external patients required for this study design as well as the accrual rate that would be needed to ensure all external patients start therapy no earlier than the first patient enrolled in the trial and no later than the last patient enrolled in the trial.

Let $n_E$ and $n_C$ be the target number of patients in the experimental and control group, respectively, in the original design (450 for each arm in IMpassion130). We simulated data for a hybrid design in which the number of experimental patients $n_E$ is fixed, but the number of control patients $n_C$ is reduced because the information can be borrowed from an external source. Let $r: 1$ be the desired randomization ratio after borrowing from an external source (experimental:control), and let $w_{down}$ be the expected downweighting factor used by the borrowing method (or implicit weighting factor for methods that don't use an explicit weight). Also, let $s$ be the total accrual rate in the trial (e.g. patients per month) assumed to be constant, and let $s_E$ and $s_C$ be the accrual rates to the experimental and control arm of the trial, respectively ($s_E + s_C = s$).

We used the following procedure to simulate data:

1. Set the number of trial controls needed in the hybrid design to $n_{C,hyb} \leftarrow n_E / r$.

2. Set the number of external (RWD) patients needed to compensate for the difference in the number of controls originally planned for the number in the hybrid design $n_{C,hyb}$ to $n_{RWD} \leftarrow (n_C - n_{C,hyb}) / w_{down}$

3. Set the accrual rates in the trial so that both arms finish accruing at the same time under the new randomization ratio $r$:

    a. $s_E \leftarrow s * r / (r + 1)$

    b. $s_C \leftarrow s / (r + 1)$

4. Determine the new length of enrollment for the study as $t_{enroll} = n_E / s_E$

5. Set the enrollment rate for the external (RWD) cohort to $s_{RWD} \leftarrow n_{RWD} / t_{enroll}$ so that accrual of external patients finishes at the same time as accrual in the trial for a fully concurrent external cohort.

6. Determine uniform accrual times as follows (rounding $n_E$, $n_{C,hyb}$, and $n_{RWD}$):

    a. Trial experimental arm: $u_{E,i} = i / s_E$, $i = 1, ..., n_E$

    b. Trial control arm: $u_{C,i} = i / s_C$, $i = 1, ..., n_{C,hyb}$

    c. External control arm: $u_{RWD,i} = i / s_{RWD}$, $i = 1, ..., n_{RWD}$

7. Generate observation times and event indicators:

    a. Generate events times as $T_i \sim Exp\,[\lambda_0 \exp(\beta_E x_{E,i} + \beta_{RWD} x_{RWD,i})]$ and censoring times

    $C_i \sim Exp[\lambda_0 \exp(\beta_E x_{E,i} + \beta_{RWD} x_{RWD,i}) \, p_{lost} / (1 - p_{lost})]$ where $\lambda_0$ is the baseline hazard, $x_{E,i}$ is an indicator for

the experimental arm, $x_{RWD, i}$ is an indicator for the external (RWD) arm, and $p_{lost}$ is the probability of being lost to follow-up.

b. Set the observed time to $Y_i = T_i \wedge C_i$ and the event indicator to $\delta_i = I[Y_i = T_i]$.

8. Censor all events after the target number of events based on calendar time, discounting external events based on how much is expected to be borrowed. Letting $t_i = Y_i - u_i$ be the calendar time for patient $i$'s observations:

   a. Set the time at which the target number of events is hit $t_{target}$ such that

   $$\sum_{i:t_i < t_{target}} \delta_i(1 - x_{RWD, i}) + \delta_i x_{RWD, i} w_{down} = d_{target}$$ where $d_{target}$ is the target number of events

   b. Censor patients at their follow-up time at the point of the clinical cutoff: For all $i$ such that $t_i > t_{target}$, set

   $Y_i \leftarrow t_{target} - u_i$ and $\delta_i = 0$

   c. Remove any patients for whom $Y_i < 0$ (these are patients that would have enrolled after the target number of events was hit, so presumably the study would have closed enrollment prior to their accrual)

For the trial data simulations we set the parameters at:

- $s = 34$ patients enrolling per month in the trial

- $\lambda_0 = 0.043$ baseline hazard in months

- $p_{lost} = 0.05$ probability of being lost to follow-up

- $w_{down} = 0.6$ expected downweighting of external controls

- $d_{target} = 655$ target events

- $\exp(\beta_{RWD}) = 0.5, 0.6, 0.7, ..., 1.8, 1.9, 2.0$ residual bias

- $\exp(\beta_E) = 0.7, 0.78, 0.85, 1$ experimental treatment effect

For each combination of HR between randomized experimental and control groups $\exp(\beta_E)$ and HR between external and randomized control groups $\exp(\beta_{RWD})$, we simulated 1,000 datasets.

## C.2 Model specifications

This section gives details on how the borrowing methods were implemented. The power prior and commensurate prior models were fit with stan (Stan Development Team 2019) using 4 chains with 10,000 iterations each, and discarding the first 5,000 iterations of each chain as the burn-in period. We then computed posterior means and quantiles to obtain point estimates and credible intervals for the log hazard ratio of the experimental treatment effect.

Table C.1 shows the tuning parameters selected for the simulations. The values in Table C.1 were determined through trial and error with the aim of achieving similar power and type I error rates for all models. Please see below for a description of the models and tuning parameters.

**Table C.1.** Tuning parameter values used in the simulations. These values were determined through trial and error with the aim of achieving at least 88% power while minimizing type I error rates.

| Model | Tuning parameter | Value |
|---|---|---|
| Test-then-pool | Significance level of test ($\alpha$) | $\alpha = 0.15$ |
| Two-step approach | Constant decay factor | $c = 8.25$ |
| Power prior | Power parameter (cohort-level downweighting factor) | $a = 0.6$ |
| Commensurate prior | Scale parameter for Half-Cauchy hyperprior | $\nu = 0.035$ |

**Test-then-pool procedure**

We conducted a log rank test comparing external to trial controls using the survival package (Therneau 2015) for R (R Core Team 2019). We pooled the external and trial data if the p-value from the test was greater than the specified significance level α, and discarded the external data otherwise. With the pooled or trial-only data (as determined by the test), we then fit an exponential regression model using the survival package (Therneau 2015).

**Two-step analog to modified power prior**

The approach described in the main text was implemented with the survival package (Therneau 2015) for R (R Core Team 2019).

**Power prior model**

Let $D_{trial}$ and $D_{RWD}$ be the set of patients in the trial and external (RWD) dataset respectively with an observed event, and let $R_{trial}$ and $R_{RWD}$ be the set of patients in the trial and RWD dataset respectively who were right censored. Also, let $f(t; \lambda)$ and $S(t; \lambda)$ be the probability density function and survival function respectively of an exponential distribution with hazard rate λ, and let $a$ be the fixed power parameter (downweighting factor). The log-likelihood for the power prior was constructed as:

$$l = \sum_{i \in D_{trial}} f(y_i; \exp(\beta_0 + x_{E,i} * \beta_1)) + \sum_{i \in R_{trial}} S(y_i; \exp(\beta_0 + x_{E,i} * \beta_1)) + a \sum_{i \in D_{RWD}} f(y_i; \exp(\beta_0)) + a \sum_{i \in R_{RWD}} S(y_i; \exp(\beta_0))$$

$$\beta_0 \propto 1$$
$$\beta_1 \propto 1$$

where $x_{E,i}$ is an indicator for the experimental group, $\beta_0$ is the log baseline hazard, and $\beta_1$ is the log hazard ratio for the experimental group.

**Commensurate prior model**

Using the same notation as above, with the addition of $\beta_{0, trial}$ to represent the log baseline hazard in the trial controls and $\beta_{0, RWD}$ to represent the log baseline hazard in the external (RWD) controls, we specified the log likelihood as:

$$l = \sum_{i \in D_{trial}} f(y_i; \exp(\beta_{0, trial} + x_{E,i} * \beta_1)) + \sum_{i \in R_{trial}} S(y_i; \exp(\beta_{0, trial} + x_{E,i} * \beta_1)) + \sum_{i \in D_{RWD}} f(y_i; \exp(\beta_{0, RWD})) + \sum_{i \in R_{RWD}} S(y_i; \exp(\beta_{0, RWD}))$$

$$\beta_{0, trial} \sim N(\beta_{0, RWD}, 1/\tau)$$
$$\tau \sim HalfCauchy(\nu)$$

where $\tau^2$ is the precision of the normal prior, and $HalfCauchy$ is a Half-Cauchy distribution with scale parameter $\nu$ and truncated to be non-negative (Gelman et al. 2008).

**No borrowing**

We fit an exponential regression model to the trial data only, excluding external data, using the survival package (Therneau 2015).

## C.3 Metrics

**Number of external events effectively borrowed**

The number of external events effectively borrowed measures the number of additional events that would have needed to occur in the control arm of the trial to obtain the same precision for the estimated experimental treatment effect in the absence of external data (Hobbs et al. 2011). For exponential data with no covariate adjustments, the variance of the treatment effect is purely a function of the number of events that have occurred. This makes it straightforward to estimate this metric for borrowing methods with an explicit downweighting factor by summing the weights for external patients who have an observed event. However, this may not be straightforward in other settings, in which the precision is also a function of the relative timing of events and covariates, or when the borrowing method does not have an explicit downweighting factor.

*Test-then-pool:* The number of events effectively borrowed is set to the number of events in the external dataset if it is pooled, and set to 0 if the external dataset is discarded.

*Two-step:* The number of events effectively borrowed is set to the product of the downweighting factor *w* and the number of events in the external arm.

*Power prior model*: The number of events effectively borrowed is set to the product of the power prior *a* and the number of events in the external arm.

*Commensurate prior model:* The number of events effectively borrowed is estimated using the approach of Hobbs et al.(Hobbs et al. 2011). In particular, in addition to the commensurate prior model we fit an exponential Bayesian model to the trial data only with diffuse priors on the regression coefficients. Let $d$ be the total number of events that occurred in the trial (excluding external patients), and let $\hat{\sigma}^2_{trial}$ and $\hat{\sigma}^2_{hyb}$ be the variances of the posterior distributions of the log hazard ratio for the experimental treatment effect in the model fit to trial data only and the commensurate prior model respectively. The effective number of events borrowed is calculated as $\max\{0, d * (\hat{\sigma}^2_{trial}/\hat{\sigma}^2_{hyb} - 1)\}$.

**Power and Type I error**

For power and type I error, we tested the one-sided null hypothesis $H_0: \beta_1 \geq 0$ against the alternative $H_1: \beta_1 < 0$ at a significance level of $\alpha = 0.025$, where $\beta_1$ is the log hazard ratio for the treatment effect. For the test-then-pool and two-step procedure, we obtained upper bounds as $\hat{U} = \hat{\beta}_1 + z_\alpha \hat{\sigma}$, where $\hat{\beta}_1$ is the estimated log hazard ratio for the treatment effect, $\hat{\sigma}$ is the standard error, and $z_\alpha = 1 - \Phi(\alpha)$ is the upper $\alpha$ quantile of a standard normal distribution. For the power prior and commensurate prior model, we set $\hat{U}$ to the upper $\alpha$ percentile of the posterior distribution of the log hazard ratio for the experimental treatment effect. For all methods, we rejected the null hypothesis if $\hat{U} < 0$.

**Mean squared error (MSE), bias, and standard deviation of number of external events effectively borrowed**

We also computed the MSE and bias of the log hazard ratio for the experimental group, and the standard deviation of the number of external events borrowed.

## C.4 Supplemental results

Figure C.1 shows the MSE, bias, and standard deviation of the number of external events effectively borrowed for all borrowing methods, and Figure C.2 shows the MSE and bias for the dynamic borrowing methods, excluding the static power prior model.

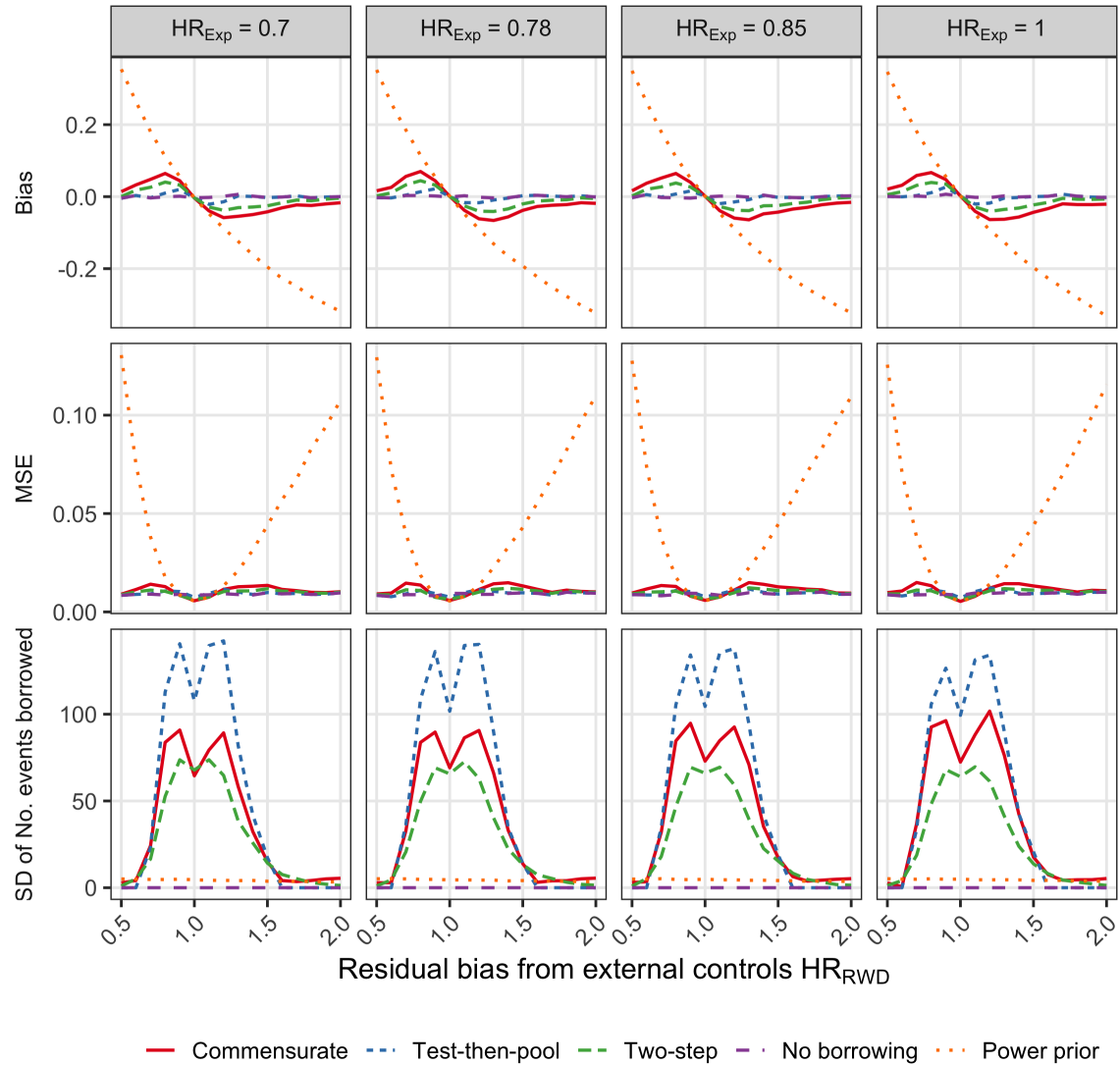

**Figure C.1.** MSE, bias, and standard deviation of the number of external events effectively borrowed for all methods

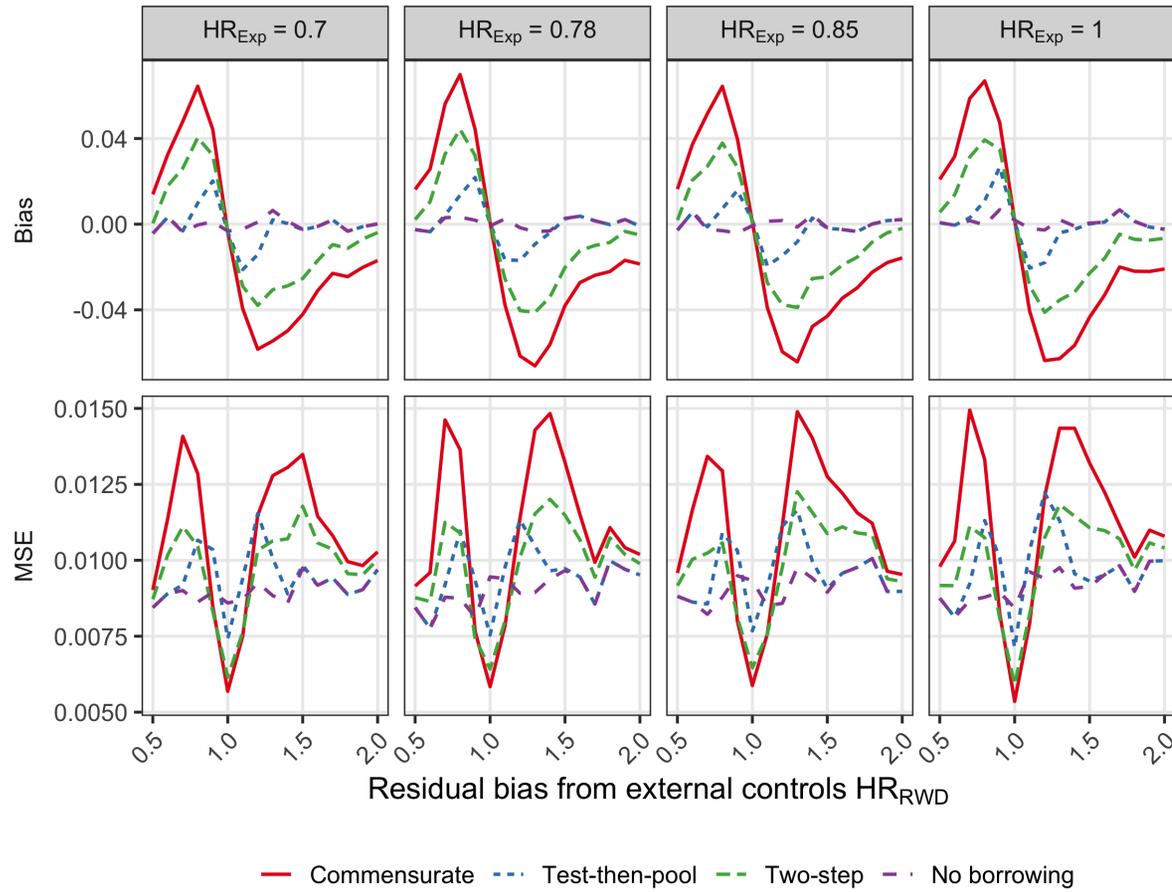

**Figure C.2.** MSE and bias. These are the same results shown in Figure C.1, but excluding the power prior model and with different y-axis scales.

Therneau, T. 2015. *A Package for Survival Analysis in S.Version 2.38*.

US Food and Drug Administration. 2020. Rare diseases: Natural history studies for drug development: Guidance for industry: Draft guidance. Accessed at https://www.fda.gov/media/122425/download on August 11, 2020